\documentclass[pra,twocolumn,footinbib,floatfix]{revtex4-2}
\usepackage{graphicx}
\usepackage{epstopdf}
\usepackage{bm,mathrsfs}
\usepackage{amsmath,bbm}
\usepackage{amsfonts,amssymb}
\usepackage{mleftright}
\usepackage{algorithm}
\usepackage{algorithmicx}
\usepackage{algpseudocode}
\usepackage{hyperref}

\renewcommand{\phi}{\varphi}

\newcommand{\bZ}{\mathbbm{Z}}

\newcommand{\bE}{\mathbbm{E}}

\newcommand{\bC}{\mathbbm{C}}

\newcommand{\1}{\mathbbm{1}}

\newcommand{\Tr}{{\rm Tr}}

\newcommand{\bq}{\begin{eqnarray}}
\newcommand{\eq}{\end{eqnarray}}

\newcommand{\be}{\begin{equation}}
\newcommand{\ee}{\end{equation}}
\newcommand{\bea}{\begin{eqnarray}}
\newcommand{\eea}{\end{eqnarray}}

\def\>{\rangle}
\def\<{\langle}

\newcommand{\ket}[1]{|#1\rangle}
\newcommand{\bra}[1]{\langle#1|}

\newcommand{\avr}[1]{\langle#1\rangle}

\newcommand{\qed}{}

\def\qed{\leavevmode\unskip\penalty9999 \hbox{}\nobreak\hfill
     \quad\hbox{\leavevmode  \hbox to.77778em{%
               \hfil\vrule   \vbox to.675em%
               {\hrule width.6em\vfil\hrule}\vrule\hfil}}
     \par\vskip3pt}
     
\begin{document}

\newtheorem{theorem}{Theorem}
\newtheorem{lemma}[theorem]{Lemma}
\newtheorem{corollary}[theorem]{Corollary}
\newtheorem{proposition}[theorem]{Proposition}
\newtheorem{definition}[theorem]{Definition}
\newtheorem{example}[theorem]{Example}
\newtheorem{conjecture}[theorem]{Conjecture}
\newenvironment{remark}{\vspace{1.5ex}\par\noindent{\it Remark:}}%
    {\hspace*{\fill}$\Box$\vspace{1.5ex}\par}

\title{Geometry of learning neural quantum states} 

\author{Chae-Yeun Park and Michael J. Kastoryano}
\affiliation{Institute for Theoretical Physics, University of Cologne, Germany}

\date{\today}
\begin{abstract}
Combining insights from machine learning and quantum Monte Carlo, the stochastic reconfiguration method with neural network Ansatz states is a promising new direction for high-precision ground state estimation of quantum many-body problems. 
Even though this method works  well in practice, little is known about the learning dynamics.
In this paper, we bring to light several hidden details of the algorithm by analyzing the learning landscape.
In particular, the spectrum of the quantum Fisher matrix of complex restricted Boltzmann machine states exhibits a universal initial dynamics, but the converged spectrum can dramatically change across a phase transition. 
In contrast to the spectral properties of the quantum Fisher matrix, the actual weights of the network at convergence do not reveal much information about the system or the dynamics. Furthermore, we identify a new measure of correlation in the state by analyzing entanglement in eigenvectors. We show that, generically, the learning landscape modes with least entanglement have largest eigenvalue, suggesting that correlations are encoded in large flat valleys of the learning landscape, favoring stable representations of the ground state.    
\end{abstract}
\maketitle

%%%%%%%%%%%%%%%%%%%%%%%%%%%%%%%%%%%%%%%%%%%%%%%%%%%%%%%%%%%%%%
%%%%%%%%%%%%%%%%%                              Introduction                            %%%%%%%%%%%%%%%%%%%%%
%%%%%%%%%%%%%%%%%%%%%%%%%%%%%%%%%%%%%%%%%%%%%%%%%%%%%%%%%%%%%%

\section{Introduction}

Recently the fields of machine learning and quantum information science have seen a lot of crossbreeding. On the one hand, a number of promising results have been obtained suggesting the potential for performing quantum or classical machine learning tasks on a quantum computer \cite{biamonte2017quantum}. 
In particular, the variational quantum eigensolver \cite{peruzzo2014variational}--perhaps the most promising quantum algorithms for first generation quantum computers--is based on the variational optimization of a cost function to be evaluated on a quantum device, providing a new playground for hybrid quantum-classical learning \cite{mcclean2016theory,kandala2017hardware}. 
However, arguably the most significant advances have been in the field of classical variational algorithms for quantum many-body systems. A number of studies have shown that machine learning inspired sampling algorithms can reach state of the art precision; including ground state energy estimation \cite{carleo2017solving,nomura2017restricted,glasser2018neural,choo2019two}, time evolution \cite{carleo2017solving,carleo2012localization}, identifying phase transitions \cite{van2017learning,carrasquilla2017machine,broecker2017machine}, and decoding quantum error correcting codes \cite{sweke2018reinforcement,Andreasson2019quantumerror} (for a recent review, see Ref.~\cite{carleo2019machine}). 

A model that has gathered a particularly large amount of attention is the complex restricted Boltzmann machine (RBM) state Ansatz with stochastic reconfiguration optimization introduced by Carleo and Troyer~\cite{carleo2017solving}. 
The authors show that ground state energy evaluations can outperform the state of the art tensor network methods on benchmark problems.

At present, however, there is lacking a theoretical underpinning for explaining why the complex RBM--or any other machine learning inspired parametrization--is a good Ansatz for describing ground states of physical Hamiltonians, or for accessing its features. 
This is sometimes referred to as the ``black box'' problem of machine learning inspired approaches, that theoretical understanding lags far behind the numerical state of the art.
In particular, it is difficult to assess and quantify the role of entanglement in these new classes of wave functions.
This is to be contrasted with the density matrix renormalization group (DMRG)~\cite{white1992density}, which was first developed as an extension of numerical renormalization group. Subsequently, it was realized that the theoretical underpinning of DMRG was the theory of tensor network states, which connect the efficiency of simulation in one dimensional systems with the amount and nature of entanglement in the spin chain. We are far form such a detailed understanding of machine learning inspired methods.

Thus it is natural that some studies have related complex RBM states to tensor network states~\cite{chen2018equivalence,collura2019descriptive}.
%where the role of entanglement is naturally built into the model. 
But these studies are mostly based on constructing abstract mappings between RBM wave functions and tensor network states, and usually  provide at best existence proofs. 

In this paper, we aim to obtain a better understanding of the learning dynamics with complex RBM wave functions by analyzing the geometry induced in parameter space. Indeed, the stochastic reconfiguration method updates the variational parameters of the wave function by gradient descent of the energy, weighted by a ``quantum Fisher matrix'', which is the quantum analog of the Fisher information matrix. The Fisher information matrix is known to be the unique Riemannian metric associated to a probability space invariant under sufficient statistics \cite{cencov2000statistical}. Hence it is the natural candidate for associating an ``information geometry'' to a statistical model. 

\begin{figure}
	\centering
	\includegraphics[width=0.9\linewidth]{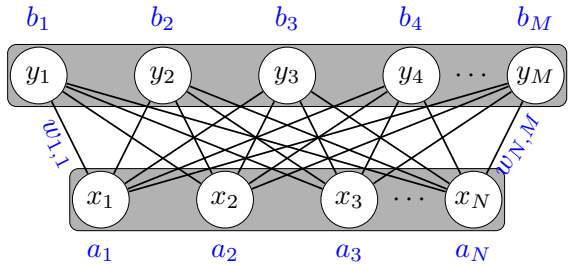}
	\caption{Complex RBM consisting of one hidden and one visible layer. Visible, hidden biases, and weights are $a\in\bC^N$, $b\in \bC^M$, and $w\in \bC^N\times\bC^M$, respectively. $x,y$ are binary vectors of length $n$ and $m$ respectively. \label{figRBM}}
\end{figure}

We analyze the spectral properties of the ``quantum Fisher matrix'' for a various lattice spin models. We argue that the information geometry provides us with clues for both the expressibility of the Ansatz state and of the underlying physics, provided the optimization converges. In particular, we identify a number of features which we believe to be universal for spin models: 

(i) The spectrum of the quantum Fisher matrix becomes singular in phases connected to a product state (in the computational basis). The singularity is more pronounced the closer one gets to the product state;

(ii) Critical phases have a smooth and extended spectrum, which is also reminiscent of image recognition models in classical machine learning;

(iii) Kinks in the spectrum reveal symmetries in the state.

(iv) The eigenvalues are exponentially decaying in value. The largest eigenvalues have eigenvectors that are dominated by first moments; i.e. they do not contain much information about correlations in the system. This feature is accentuated the sharper the spectrum profile of the quantum Fisher matrix.

The above insight was extracted from extensive numerical data calculated using quantum spin Hamiltonians such as transverse field Ising and Heisenberg spin-XXZ models as well as coherent Gibbs states for the two dimensional classical Ising model. Various Monte Carlo sampling strategies were used to optimize the results on large system sizes.

Importantly, we observe that the bare values of the variational parameters reveal very little information about the physical properties of the system, contrary to what is often claimed that ``activations indicate regions of activity in the underlying data''. We take this as evidence that there are many equivalent representations of the states in the vicinity of the ground state, suggesting that  the optimizer preferentially choses robust representation of the ground state. Robustness of the Monte Carlo methods might be related to the generalization property in supervised learning. 
Our study shows that the spectrum of the quantum Fisher matrix can be an essential diagnostic tool for further exploration with complex RBM wave functions as well as with other machine learning inspired wave functions.

\subsection{Complex RBM and optimization by stochastic reconfiguration}
The complex restricted boltzmann machine (RBM) neural network quantum state specifies the amplitudes of a wave function $\ket{\psi_\theta}=\sum_x \psi_\theta(x)\ket{x}$ in some chosen computational basis $\{\ket{x}\}$ by the exponential family:
\be \psi_\theta(x) = \sum_{y} e^{a\cdot x+b\cdot y +x^T w y}/\sqrt{Z},\ee
where the vectors $\{a,b\}$ and the matrix $w$ contain complex parameters to be varied in the optimization, and $y$ is a binary vector indexing ``hidden'' units. $Z=\sum_x |\psi_\theta(x)|^2$ is a constant guaranteeing normalization of the state $\psi$.  The complex RBM can be visualized as a binary graph $(V,E)$ between the visible nodes $x$ and the hidden nodes $y$ [see Fig.~\ref{figRBM}]. To each edge $e\in E$ we associate a variational parameter $w_e$, and at each vertex $v\in V$ we associate a bias weight $a$ or $b$ to a visible ($x$) or hidden ($y$) binary degree of freedom. We will often express the variational parameters as a concatenated vector labeled $\theta=(a,b,{\rm vec}(w))$. 
For classical RBMs, the normalization constant is the partition function of a joint probability distribution on the hidden and visible units. This is generally not true in the complex case.

The goal of variational Monte Carlo is  to find the optimal parameters $\theta$ that minimize the energy of a given Hamiltonian in the state $\ket{\psi_\theta}$. The standard approach would be to use gradient descent, but this performs very poorly for spin Hamiltonians, as the updates tend to get stuck oscillating back and forth along  steep wells of the energy landscape rather than falling down the more shallow directions. 
The stochastic reconfiguration (SR) method \cite{sorella1998green,sorella2001generalized} for energy minimization is derived as a second order iterative approximation to the imaginary time ground state projection method (see Appendix \ref{sec:app_sr} for a self contained derivation). In SR, the parameters of the Ansatz wave function are iteratively updated as 

\be \theta \rightarrow \theta - \eta S^{-1}\nabla_\theta \avr{H}, \label{eq:SR_update} \ee
where $\eta$ is a constant specifying the rate of learning. The second order effects which take curvature into account are determined by the matrix
\be S_{\alpha\beta}=\avr{O^\dag_\alpha O_\beta}-\avr{O^\dag_\alpha}\avr{ O_\beta},\label{eqn:qFisher}\ee
of the diagonal operators $O_\alpha$, with $\alpha\in\theta$, which act for instance as
\be O_{w_{ij}}\ket{x}=\frac{\partial \log \psi_\theta(x)}{\partial w_{ij}}\ket{x},\ee
in the computational basis $\{ x\}$. 
We will call the matrix $S$ the \textit{quantum Fisher matrix}, because of its connection with information geometry as discussed in detail in the next section.  The quantum Fisher matrix can be reformulated as a classical covariance matrix of the operators $O_\alpha,O_\beta$,
\be S_{\alpha\beta}=\bE[O^\dag_\alpha O_\beta]-\bE[O^\dag_\alpha]\bE[ O_\beta]\label{eq:S_elt},\ee
and similarly
\be \partial_{\alpha} \avr{H}=\bE[O_\alpha H_{loc}]-\bE[O_\alpha]\bE[ H_{loc}],\label{eq:Egrad}\ee
where $\bE[A]=\sum_{x} A(x) |\psi_\theta(x)|^2$ is the classical expectation of operator $A$ in the state $|\psi_\theta(x)|^2$, and 
\be H_{loc}(x)=\frac{\avr{x|H|\psi_\theta}}{\avr{x|\psi_\theta}}\ee
is called the local energy. 

For the RBM Ansatz, the diagonal operators $ O_\alpha$ take on the simple form:
\begin{align}
	O_{a_i}(x) &= x_i\\
	O_{b_j}(x) &= \tanh \chi_j(x)\\
	O_{w_{ij}}(x) &= x_i \tanh \chi_j(x) 
\end{align}
where $\chi_j(x)=b_j + \sum_i w_{ij} x_i$, and indices $i$ run over $[1, \cdots, N]$ visible vertices and $j$ run over  $[1, \cdots, M]$ hidden vertices.
Thus the size of the quantum Fisher matrix is $N+M+NM$.

The SR method is computationally efficient when the following are true:
\begin{enumerate}
\item The operators $O_\alpha(x)$ and $H_{loc}(x)$ can be computed efficiently for every point $x$.
\item The probability distribution $|\psi_\theta(x)|^2$ can be sampled from for any values of $\theta$; meaning that any single Monte Carlo update can be computed efficiently. In practice we require that each Monte Carlo update is independent of system size; i.e. updates are local. 
\item The sampling procedure converges rapidly (in sub-polynomial time) to the desired state $|\psi_\theta(x)|^2$.
\end{enumerate}

The complex RBM Ansatz guarantees that (1) and (2) hold whenever the number of hidden units is a constant multiple of the visible units. However, like essentially any sampling algorithm, provably guaranteeing (3) seems nearly impossible in any practically relevant problem. However, experience has shown that convergence often is rapid in practice, or can be curtailed, whenever one steers clear of frustration or the Fermionic sign problem. It is worth pointing out, though, that convergence of the sampler can depend sensitively on the chosen basis and the initial state, as evidenced in Sec. \ref{sec:XXZ}.

\subsection{Natural gradient and SR}
The SR method~\cite{sorella1998green,sorella2001generalized} can be interpreted geometrically~\cite{mazzola2012finite}, which makes a direct connection to
Amari's natural gradient optimization~\cite{amari1998natural}.
Plain vanilla gradient descent optimizes a multivariate function $L(\theta)$ by updating the parameters in the direction of steepest descent:
\be \theta \rightarrow \theta-\eta\nabla_\theta L(\theta),\ee
at a certain rate $\eta$. 

In systems where the landscape of the  function $L(\theta)$ is very steep in certain directions and shallow in others, convergence can be very slow as the updates fluctuate back and forth in a deep valley, but take a long time to ``drift'' down a shallow one. The natural gradient method proposes to update the parameters according to the natural (Riemannian) geometric structure of the information space, so that the landscape is made locally euclidean before the update.  
Suppose the coordinate space is a curved manifold in the sense that the infinitesimal square length is given by the quadratic form

\be {\rm d} s^2=\sum_{\alpha\beta} g_{\alpha\beta}(\theta){\rm d}\theta_\alpha{\rm d}\theta_\beta,\ee
where the matrix $g(\theta)$ is the Riemannian metric tensor. Amari showed that the steepest descent direction of the function $L(\theta)$ in the Riemannian space is given by 
\be -\tilde{\nabla}(\theta)=-g^{-1}(\theta)\nabla L(\theta).\ee

The action of the inverse of $g$ can be heuristically understood as ``flattening'' out the space locally. 
For general optimization problems, the Hessian is a natural choice for $g(\theta)$, as it reproduces Newton's second order method. 
In machine learning applications, and with RBMs in particular,  the Hessian is hard to construct from  sampling. It also appears to be attracted to saddle points~\cite{dauphin2014identifying}.

When the parameter space in question is naturally associated with a classical probability distribution, the ``natural'' geometry is chosen to be the Fisher information matrix as it is the unique metric that is invariant under sufficient statistics \cite{cencov2000statistical}. 
For pure parametrized quantum states, the natural Riemannian metric is derived from the Fubini-Study distance: 
\begin{align}
	\gamma(\psi, \phi) = \arccos \sqrt{\frac{\<\psi|\phi\>\<\phi|\psi\>}{\<\psi|\psi\>\<\phi|\phi\>}}.
\end{align}
Infinitesimal distances are given by:
\begin{align}
	ds^2 = \gamma(\psi,\psi+\delta \psi)^2 = \frac{\<\delta\psi|\delta\psi\>}{\<\psi|\psi\>} - \frac{\<\delta \psi|\psi\>}{\<\psi|\psi \>}\frac{\<\psi|\delta \psi\>}{\<\psi|\psi \>}
\end{align}
which reproduces the quantum Fisher matrix for parametrization $\theta$ as $ds^2 = \sum_{\alpha\beta}S_{\alpha\beta} d \theta_\alpha^* d\theta_\beta$.

In particular, when the  wave function is positive in a given computational basis, the quantum state can be written as $\ket{\psi} = \sum_{x} \sqrt{p_\theta (x)} \ket{x}$, and the quantum Fisher matrix is
\begin{align}
	S_{\alpha\beta} &= \frac{1}{4}\bigl< \frac{\partial \log p_\theta (x)}{\partial \theta_\alpha} \frac{\partial \log p_\theta (x)}{\partial \theta_\beta} \bigr> \nonumber \\
	&\quad - \frac{1}{4}\bigl< \frac{\partial \log p_\theta (x)}{\partial \theta_\alpha}\bigr>\bigl< \frac{\partial \log p_\theta (x)}{\partial \theta_\beta}\bigr> \\
		   &= \frac{1}{4} \mathcal{F}_{\alpha\beta}
\end{align}
where $\< A \> = \mathbb{E}[A]$ and $\mathcal{F}$ is the Fisher information matrix associated to the probability distribution $p_\theta(x)$.
Thus, the SR method reproduces the natural gradient method for positive wave functions. For this reason, we will be calling the $S$ matrix associated to a pure quantum state the \textit{quantum Fisher matrix}.

\subsection{Spectral analysis of the quantum Fisher matrix}

In this paper, we will argue that spectral properties of the quantum Fisher matrix reveal essential information about the physical properties of the system under study as well as  the dynamics of optimization.

The quantum Fisher matrix is positive semi-definite, implying that its spectrum is real and there exists a set of orthonormal eigenvectors. The magnitude of an eigenvalue determines how steep the learning landscape is in that particular direction. The spectrum will generically be sloppy \cite{sloppy,machta}, with a spectral function bounded above by a decaying exponential. 

It is often argued in the machine learning community that gradient descent algorithms favor regions in parameters space where most eigenvalues are close to zero~\cite{sagun2016eigenvalues,papyan2018full}. This implies that at convergence, most directions in the landscape are nearly flat, suggesting that nearby points in parameter space encode much of the same physical properties. In classical supervised learning, the flatness of the landscape has been associated with the ``generalization'' ability of the learned model~\cite{hochreiter1997flat}; in the physics setting we interpret it to mean that the representation is robust. 

Because of the bipartite graph structure of the RBM Ansatz, it is natural to talk about correlations between the visible and hidden units. The quantum Fisher matrix is a square $(N+M +NM)$ matrix, with the first two blocks corresponding to the biases $a,b$, and the third block corresponds to the weights matrix $w$. The main $w$ block describes the orientations in parameter space that can affect correlations in the model.  
We will see later that eigenvectors associated to eigenvalues of large magnitude are typically close to a product state between the visible and hidden part, meaning that they mostly just affect the first moments of the spin variables. 

To measure correlations in the eigenvectors $\{\psi_\alpha\}$, we truncate the first two blocks of the eigenvectors associated with the biases, and renormalize the ``$w$'' part to have Hilbert Schmidt norm 1. We then calculate the entanglement in the eigenstate $\psi^w_\alpha$:
\begin{equation}
{\rm Ent}(\psi_\alpha)=S(\Tr_h[\psi^w_\alpha]),\label{eqn:entev}
\end{equation}
where $\Tr_h$ is the partial trace over the hidden layer, and $S(\cdot)$ is the von Neumann entropy of the reduced density matrix.

\begin{figure*}[t]
	\centering
	\includegraphics[width=1.0\linewidth]{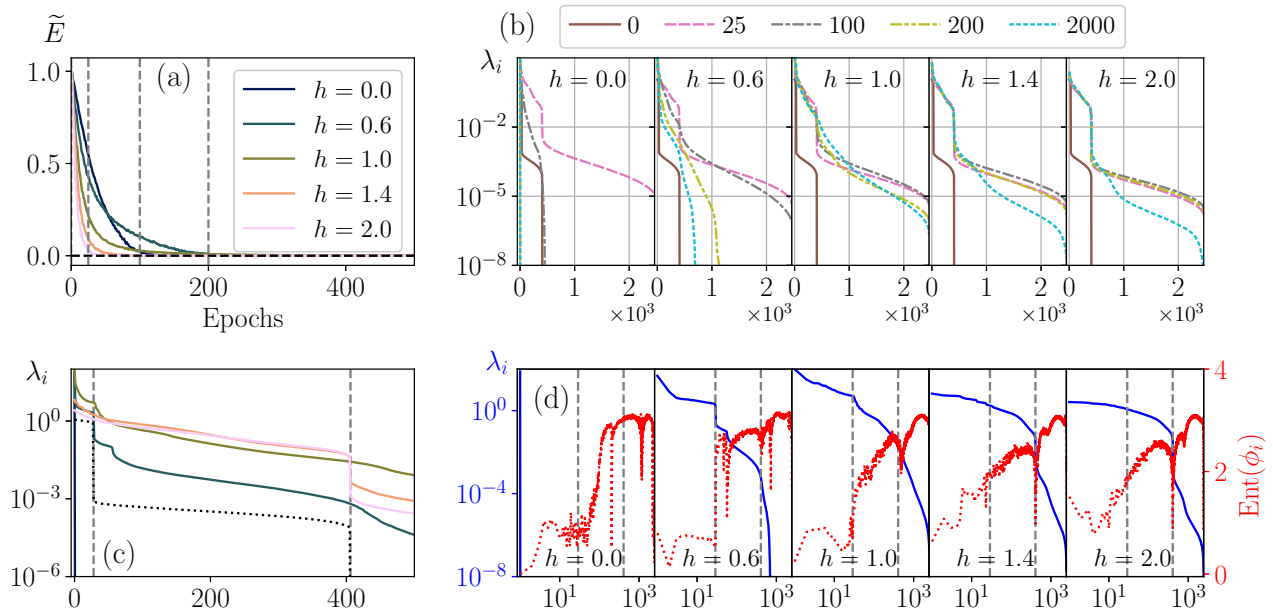}
	\caption{\label{fig:TFI_evals} 
		Transverse field Ising model, variational ground state energy optimization using the SR: (a) Rescaled energy as a function of epochs for different values of $h\in[0.0, 0.6, 1.0, 1.4, 2.0]$ (from darkest to lightest). The energy is rescaled to have $0$ at the exact ground state energy and $1$ at initialization. (b) Ordered eigenvalues of the quantum Fisher matrix [Eq.~\eqref{eqn:qFisher}] at epochs $0$ (solid), $25$ (dashed), $100$ (dot-dashed), $200$ (dot-dot-dashed), and $2000$ (dotted). Results from $h=0.0$ (the leftmost) to $h=2.0$ (the rightmost) are shown in each subplot.
		The spectrum exhibits universal behavior for the first $\sim25$ epochs. After that, the eigenvalues slowly approach a model dependent final profile (see main text). (c) The 500 largest eigenvalues after convergence for different values of $h$ as well as for randomly initialized RBM (black dotted curve). Color coding is the same as in (a). The two vertical gray dashed lines indicate $N=28$ and $N(N+1)/2=406$. (d) Spectrum (blue solid) and entanglement in the eigenvectors (red dotted) on log-log scale. The eigenvectors corresponding to the dominant eigenvalues have significantly reduced entanglement, especially in the ferromagnetic phase. Hyper-parameters $\eta=0.01$ and $\epsilon=0.001$ are used.
	}
\end{figure*}

\section{Results}\label{sec:results}
In this section, we analyze the spectral properties of the quantum Fisher matrix during the learning process of finding the ground state of the transverse field Ising (TFI) model.  The TFI Hamiltonian is given by
\begin{align}
	H =  - \sum_{i=1}^N \sigma^{i}_z \sigma^{i+1}_z - h \sum_{i=1}^N \sigma^{i}_x  \label{eq:hamTFI}
\end{align}
where $\pmb{\sigma}^{i} = \{\sigma^{i}_x,\sigma^i_y,\sigma^i_z\}$ are Pauli spin operators, and $h$ is the external field.
The system has $\mathbb{Z}_2$ symmetry ($\sigma^i_z \rightarrow -\sigma^{i}_z$) which is explicitly broken for $h < 1$ in the thermodynamic limit ($N \rightarrow \infty$).
A second order phase transition occurs at $h=1$. At zero external field the model has two degenerate ground states $|0\rangle^{\otimes N}$ and $|1\rangle^{\otimes N}$, whereas in the limit of $h\rightarrow \infty$ the ground state is unique, given by $|+\rangle^{\otimes N}$.

We trained the RBM for this model with $N=28$ and $\alpha=M/N=3$.
The spectral properties of the quantum Fisher matrix, as well as the energy during the learning process obtained from the simulation  are plotted in Fig.~\ref{fig:TFI_evals} (details of the simulation are described in Appendix~\ref{sec:app_numeric}).
Figure~\ref{fig:TFI_evals}(a) confirms that the optimization procedure successfully finds the ground state for all values of $h$, albeit at different speeds.  
The quantum Fisher matrix is constructed approximately by Monte-Carlo sampling  and its full  spectrum is  evaluated every 5 epochs during learning.
The eigenvalues at some representative epochs are plotted in decreasing order in Fig.~\ref{fig:TFI_evals}(b). 

The dynamics of the learning process proceeds in two distinct stages.  
The first stage is observed at the very beginning of the learning, lasting for roughly $25$ epochs~\footnote{The duration of the first phase appears to depend on the hyperparameters (learning rate, regularization), but not on the system size.}, and is the same for all values of $h$. The initial shape of the spectrum has two sharp drops located at $N$ and $N(N+1)/2$ [see Fig.~\ref{fig:TFI_evals}(c)]. This is a consequence of the random initialization with small weights. An analytic justification of this behavior is provided in Appendix~\ref{sec:random_RBM}. The spectrum then gets pushed up until approximately the $25$'th epoch, revealing that more and more dimensions in the information space become relevant.  

The second stage of learning then slowly transforms the distribution to that of the final converged state.
We observe that the spectrum falls off very sharply (exponentially) in all cases examined [Fig.~\ref{fig:TFI_evals}(b)], but the exact spectral profile depends strongly on the details of the model, yet not on the system size or on the specific values of the learned weights (see Appendix C for an in depth discussion). We take this as evidence that the learned state not only minimizes the energy, but also closely matches the actual ground state of the model (that we also checked using the spin-spin correlation functions). The behavior of the spectrum of the quantum Fisher matrix for each phase of the TFI model is discussed in the next subsection.

\subsection{Phases of the TFI model}
\paragraph{The ferromagnetic phase $(h< 1.0)$.} Let us start by considering the extreme case with $h=0.0$.
The quantum Fisher matrix after convergence becomes a pure state up to numerical precision.  
The singularity of the quantum Fisher matrix in this case can be explained from the properties of the ground state:
When $h=0.0$, the Hamiltonian Eq.~\eqref{eq:hamTFI} has two ground states $|0\>^{\otimes N}$ and $|1\>^{\otimes N}$.
We first note that the optimization consistently found a solution with $a \approx 0$ and $b \approx 0$, leading to a $\mathbb{Z}_2$ symmetric state.
Let us therefore assume that the solution we have exactly describes the $\mathbb{Z}_2$ symmetric ground state; i.e. $a = b = 0$.
Then the ground state is $|0\>^{\otimes N} + |1\>^{\otimes N}$ leading to an RBM representation $|\psi_\theta(x)|^2 = 1/2$ for $x = x_0$ or $x = -x_0$ where $x_0 = [1 \cdots 1]$, and zero otherwise.

Moreover, we have $O(x_0) = [x_0, y_0, x_0\otimes y_0]$ and $O(-x_0) = [-x_0, -y_0, x_0 \otimes y_0]$ where $y_0  := [\tanh \chi_1(x_0), \cdots, \tanh \chi_m(x_0)]$.
This gives 
\begin{align}
\mathbb{E}[O] &= \begin{pmatrix} 0 & 0 & x_0 \otimes y_0 \end{pmatrix} \\
\mathbb{E}[O^\dagger O] &= \frac{1}{2} \bigl[ O(x_0)^\dagger O(x_0) +  O(-x_0)^\dagger O(-x_0) \bigr ] \nonumber \\
			&= \begin{pmatrix}
				x_0^\dagger x_0 & x_0^\dagger y_0 & 0 \\
				y_0^\dagger x_0 & y_0^\dagger y_0 & 0 \\
				0 & 0 & (x_0\otimes y_0)^\dagger (x_0 \otimes y_0) 
			\end{pmatrix}.
\end{align}
Thus, the quantum Fisher matrix is 
\begin{align}
	S &= \begin{pmatrix}
				x_0^\dagger x_0 & x_0^\dagger y_0 & 0 \\
				y_0^\dagger x_0 & y_0^\dagger y_0 & 0 \\
				0 & 0 & 0
				\end{pmatrix} \nonumber \\
	  &= \begin{pmatrix} x_0 & y_0 & 0 \end{pmatrix}^\dagger \begin{pmatrix}x_0 & y_0 & 0 \end{pmatrix},
\end{align}
which is rank $1$. 
We note that the above argument does not depend on the details of the weights $w$, rather only on its magnitude $|w|$, so that any set of RBM weights that accurately model the ground state will exhibit the same behavior. The SR optimization typically favors small weights. 

As the external field $h$ increases, the number of terms of the ground state  in the computation basis increases, thus we also expect that rank of  $S$ to increase as $\mathbb{E}[{O}^\dagger {O}] = \sum_{x} |\psi_\theta(x)|^2 {O}(x)^\dagger {O}(x)$.
This is consistent with the results from our numerical data in Fig.~\ref{fig:TFI_evals}(b). Importantly, rank deficiency is observed throughout the ferromagnetic phase, albeit much more pronounced in the vicinity of $h=0$. We interpret this behavior as a signature that the phase is connected to a product state in the physical basis. For values of $h$ close to one, the rank deficiency can only be seen for at large system sizes, and after many training epochs.

\paragraph{The critical point ($h = 1.0$)}
At the critical point, the distribution of eigenvalues after convergence is smooth, and decreasing exponentially. This behavior is also seen in many classical image processing tasks in machine learning \cite{papyan2018full,grosse2015scaling}, suggesting that it might be signature of (critical) long range order.  
Indeed, each element of the quantum Fisher matrix can be expanded in terms of correlation functions, all of which are sizable in the critical case. This eigenvalue distribution is characteristic of ``Sloppy model universality'', which has been shown to reflect systems with certain forms of scale invariance \cite{sloppy,machta}, further corroborating the claim. We will see in section \ref{furtherEx} that this behavior is seen in many other systems and reveals that the RBM is fine tuning a solution with the help of a large number of hidden units.

\paragraph{The paramagnetic phase ($h > 1.0$)}
In this case, we see that the energy converges rapidly and the eigenvalues almost do not change  after the initial learning stage. 
In particular  the second jump in the spectrum of the initial random RBM survives until the end.
When $h=2.0$, the jump is located at $N+N(N-1)/2 = 406$, revealing that the quantum Fisher matrix has no support on the anti-symmetric subspace (see Appendix B).
Precisely, the 406th eigenvalue has magnitude $\approx 4.08\times 10^{-2}$ and the next one has magnitude $\approx 1.38\times 10^{-3}$ in our numerical data.

To understand the stepwise behavior, we first focus on the randomly initialized RBM case; i.e. at epoch $0$.
As we initialize the parameters of the RBM with small random Gaussian values (sampled from $\mathcal{N}(0,\sigma^2)$ where $\sigma\sim 10^{-2}$), the classical probability distribution $|\psi_\theta (x)|^2$ would be similar to the case when all parameters are zero. When $a=b=w =0$, the RBM gives $|\psi_\theta(x)|^2 = \1/2^N$, i.e. the identity distribution. 
We can then perturbatively expand the quantum Fisher matrix in terms of the parameters. 
The derivation up to $O(\sigma^3)$  is given in Appendix~\ref{sec:random_RBM}.
Our derivation gives $N$ eigenvalues of $O(1)$ associated with the visible biases block of the matrix  and $N(N-1)/2$ eigenvalues of order $O(\sigma^2)$ in the weights block of the quantum Fisher matrix.
This explains the first and the second jumps in the eigenvalue distribution of the random RBM.

The randomly initialized RBM also hints at the fact that the quantum Fisher matrix throughout the paramagnetic phase strongly retains properties of the $h\gg 1$ limit with product state $\ket{+}^{N}$.
We can compare the spectra of the quantum Fisher matrix for $h=2.0$ and the randomly initialized case in Fig.~\ref{fig:TFI_evals}(c). It shows that the second step is preserved but the first step disappears. 
This is because the first step depends on the details of weights but the second one is the consequence of the symmetry. 
We made detailed comparison between the quantum Fisher matrix for the paramagnetic phase and randomly initialized RBM in Appendix~\ref{sec:more_fisher}. We there show that the converged matrix has larger diagonal elements in the $w$ part of the matrix than the random RBM case which also support eigenvalues between $N$ to $N(N+1)/2$.

Throughout the phase diagram of the TFI, the spectrum of the quantum Fisher matrix at convergence has two special points at $N$ and at $N(N+1)/2$, as seen in Fig.~\ref{fig:TFI_evals}(c). The location of these points is independent of the number of hidden units, suggesting that they originate from the $\bZ_2$ nature of the physical system, and the overall bipartite structure of the RBM, rather than any details of the RBM graph.

\subsection{Eigenvectors}
Above we have argued the eigenvalues of the quantum Fisher matrix reveal signatures of the phase of matter being simulated. We now ask whether the eigenvectors can teach us anything about how correlations are conveyed in the learning landscape. In particular, since the complex RBM is constructed from a bipartite graph with no connections among the hidden and visible units, we know that all correlations have to be mediated by weights. Entanglement in the information manifold is therefore completely contained in the weights block of the Fisher matrix. 

In Fig.~\ref{fig:TFI_evals}(d), we plot the entanglement between the visible and hidden units of the $w$ part of each eigenvector [see Eq.~(\ref{eqn:entev})]. We observe that the first $N$ eigenvectors have very little entanglement when $0\leq h \leq 1$. This suggests that the directions of largest curvature are almost exclusively associated with the biases, or first moments, of the distribution. Note that this does not imply that the values of the $w$ weights are small, as representations of the first moments are distributed over the biases and the weights. Rather it is a reminder that the actual values of the weights of the network reveal little information of the correlations in the system, as is manifest in Fig.~\ref{fig:TFI_Smat} of Appendix C. This behavior is less pronounced for $h>1$ as the quantum Fisher matrix behaves more like a random matrix whose eigenvectors are expected to have a more homogeneous amounts of entanglement.  

The entanglement increases in the bulk of the spectrum. Interestingly, this means that the directions in parameter space that encode information about correlations are typically dense, smooth and flat. In the context of classical ML, these properties are akin to good generalization ability of the learning models, whereas in the present physics context, we interpret it to meant that the algorithm preferentially learns stable configurations; where changes (even large) in most directions in configuration space will not affect the physically observable properties of the system. Similar conclusions have been alluded to in the context of sloppy models universality in statistical mechanics \cite{sloppy,machta}.

\begin{figure*}[t]
	\centering
	\includegraphics[width=1.0\linewidth]{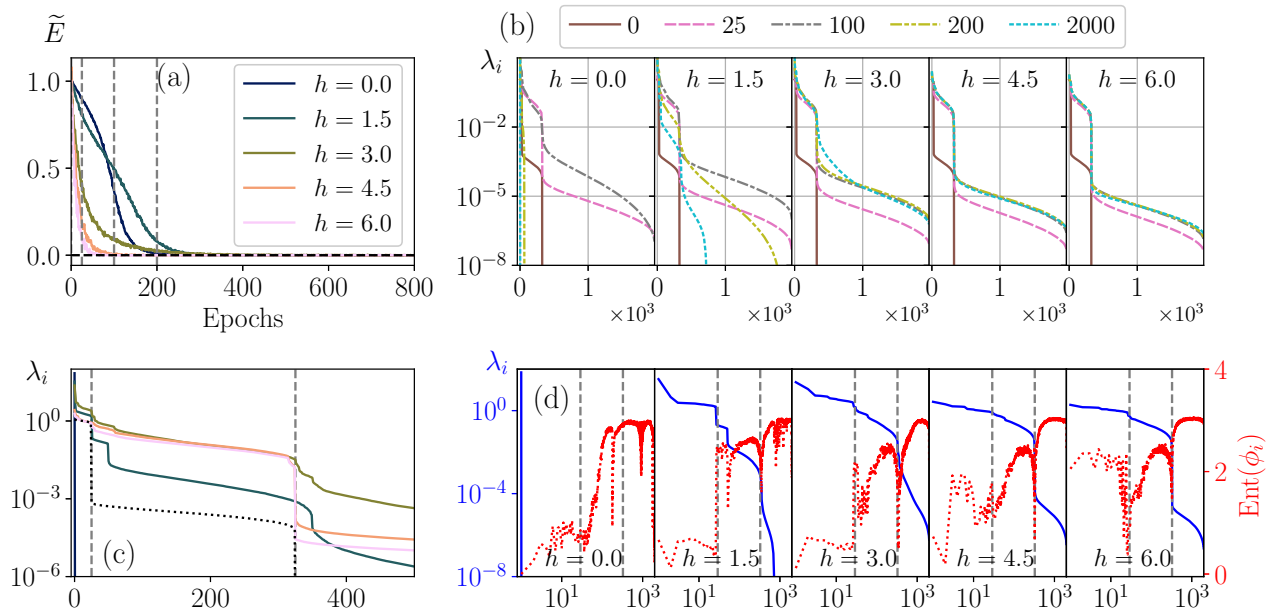}
	\caption{\label{fig:transverse_ising2d}
		Two dimensional transverse field Ising model in $5\times5$ lattice: (a) Rescaled energy as a function of epochs for $h=[0.0, 1.5, 3.0, 4.5, 6.0]$ (from darkest to lightest). (b) Ordered eigenvalues of the quantum Fisher matrix [Eq.~\eqref{eqn:qFisher}] at epochs $0$ (solid), $25$ (dashed), $100$ (dot-dashed), $200$ (dot-dot-dashed), and $2000$ (dotted). The results from $h=0.0$ (leftmost) to $h=6.0$ (rightmost) are shown in each subplot. (c) The 500 largest eigenvalues after convergence and for randomly initialized RBM (black dotted curve). The same color coding as in (a) is used. Two gray lines indicate $N=25$ and $N(N+1)/2=325$. (d) Spectrum (blue solid) and entanglement in the eigenvectors (red dotted) on log-log scale.  
	Hyper-parameters $\eta = 0.002$ and $\epsilon = 0.001$ are used.
	}
\end{figure*}

\subsection{Predictions}\label{predictions}
From the spectral analysis of the quantum Fisher matrix for the transverse field Ising model, we make the following predictions, which we expect to hold more generally for ferromagnetic quantum spin models: 

\begin{enumerate}
\item The spectral profile is universal within a phase of the model, and is only weakly dependent on system size away from phase transition points. The spectrum of the  quantum Fisher matrix is therefore a good indicator of the existence of a phase transition if it is possible to find two points in phase space with vastly different spectral profiles. 
\item The first $N$ eigenvectors are  close to product states, and hence do not encode correlations in the system. They mostly pertain to first moments of the distribution. 
\item A rank deficient  quantum Fisher matrix is evidence that the state is in a phase connected to a product state in the chosen computational basis. A smoothly decaying spectrum is a sign that the system contains a lot of correlation; often a critical phase with polynomial decaying correlation functions. 
\item Kinks in the spectrum reveal symmetries in the model. In the case of the TFI, the persistent kink at $N(N+1)/2$ is a sign that the symmetric and anti-symmetric subspaces are strictly separated everywhere except at the critical point. 
\end{enumerate}

\section{Further experiments }
In this section, we study three further models to test whether the predictions made in Sec.~\ref{predictions} extend to more general spin systems. 
The first model is the two dimensional transverse field model which is not known to be exactly solvable. 
The second is the coherent Gibbs state, whose quantum Fisher matrix is evaluated exactly without having recourse to learning. 
These two model exhibit $Z_2$ symmetry breaking as in the one dimensional transverse Ising model that we studied above. 
For these models, we find the similar quantitative behaviors of the Fisher matrix which strongly suggest the universality of our predictions. 
Our last example is the XXZ model, where we explore the Fisher matrix in all three phases.

\subsection{Two dimensional transverse Ising model}
We consider the Hamiltonian defined in a $L \times L$ two-dimensional lattice given as
\begin{align}
	H = - \sum_{\<i,j\>} \sigma^i_z \sigma^j -h \sum_i \sigma^i_x
\end{align}
where the first summation is over all nearest neighbors $\<i,j\>$ of the lattice. 
The essential physics is the same as the one dimensional model, i.e. the system is in the ferromagnetic phase when $h < h_c$ and paramagnetic phase when $h>h_c$. However, the critical point $h_c$ is only approximately known $\approx 3.00 \pm 0.05$ as the system is not exactly solvable in this case~\cite{hamer2000finite,suzuki2012quantum}.

For the system size $L=5$ that we can directly compare with the exact diagonalization, we simulated the system and plot the normalized energy and the spectral profiles of the Fisher matrix in Fig.~\ref{fig:transverse_ising2d} (a,b).
We clearly see the rank deficiency for $h = 0.0$ and $1.5$, smooth spectrum at $h\approx h_c$, and kinks when $h=4.5$ and $6.0$ which confirms the universality of our predictions.
In addition, Fig.~\ref{fig:transverse_ising2d}(c) verifies that the kinks are located at $N(N+1)/2$ and Fig.~\ref{fig:transverse_ising2d}(d) indicates low entanglement between hidden and visible layers in leading eigenvectors.

\subsection{Coherent Gibbs state of the two dimensional classical Ising model}\label{furtherEx}
We next consider the RBM representation of the coherent Gibbs state of the two dimensional classical Ising model. Recall the classical Ising model 
\begin{align}
	H(x) = - J \sum_{\<i,j\>} x_i x_j
\end{align}
where $x$ is the configuration of the spin and $\<i,j\>$ are nearest neighbors  on a two dimensional lattice. For convenience, we set $J = 1$.
We consider a system in thermal equilibrium with inverse temperature $\beta = 1/T$.
At high temperature $\beta < \beta_c$, the system exhibits a disordered paramagnetic phase characterized by zero magnetization $\<x\> = 0$ ,
whereas it shows a $\mathbb{Z}_2$ symmetry broken ferromagnetic phase with non-zero magnetization at sufficiently low temperature $\beta > \beta_c$~\cite{baxter1982}.
The phase transition takes place at $\beta = \beta_c \approx 0.44$ in the thermodynamic limit and is  second-order.
We thus have polynomial decay of the correlation function $\<x_i x_j\>_c \sim 1/{\rm dist}(i,j)^\alpha$ at the critical point.

The coherent Gibbs state for the model with  inverse temperature $\beta$ is given by
\begin{align}
	\ket{\phi(\beta)} = \sum_{\{x\}} \frac{e^{-\beta H(x)/2}}{\sqrt{Z}} \ket{x}
\end{align}
in a chosen computational basis $\{x\}$ and $Z = \sum_{\{x\}} e^{-\beta H(x)}$ is the normalization factor which is the same as the partition function of the classical model.
A key observation  is that correlation functions of spin-$z$ operators are exactly the same as that of the classical model, i.e. $\< \phi(\beta) |\sigma^i_z \sigma^j_z |\phi(\beta) \> = \<x_i x_j\>_{x\sim p(x)}$ where $p(x) = e^{-\beta H(x)}/Z$ is the Boltzmann distribution.
Thus we also have polynomially decaying quantum correlation functions for this state at $\beta = \beta_c$.
We also note that even though this state is artificially constructed, the state is a ground state of a Hamiltonian that is local in a two-dimensional lattice~\cite{perez2007peps}.

\begin{figure}[t]
	\centering
	\includegraphics[width=1.0\linewidth]{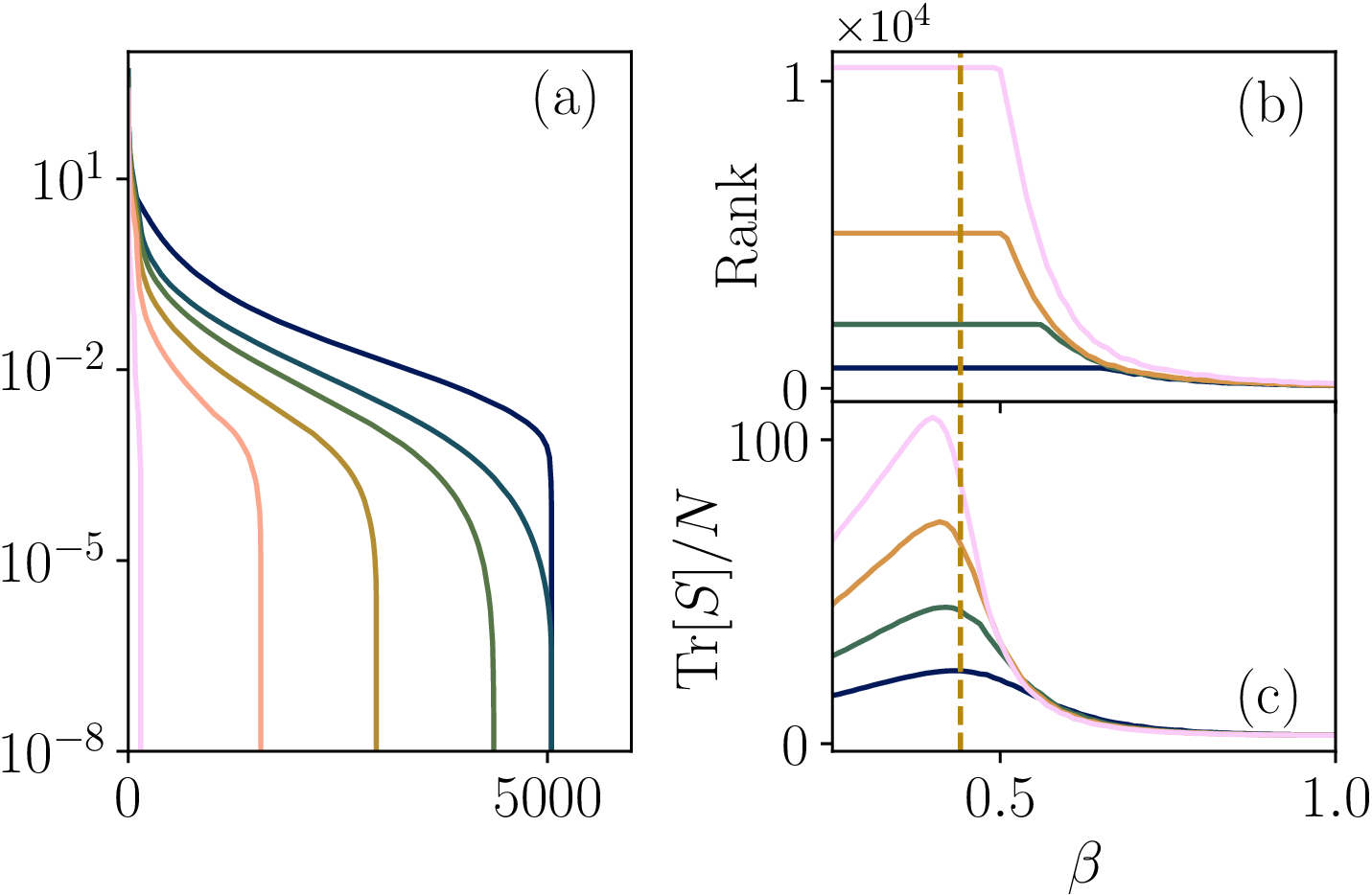}
	\caption{\label{fig:coherent_ising2d}
		(a) Eigenvalue distributions of the quantum Fisher matrix for coherent Gibbs states of two dimensional classical Ising model. The inverse temperature $\beta \in [0.10, 0.50, 0.52, 0.55, 0.6, 0.9]$ (from darkest to lightest) are used. We used $L \times L$ lattice with $L=10$, so $N=100$.
		The number of hidden units $M$ is given by the number of edges in the graph which is $180$ (open boundary condition is used). The step is exactly located at $N(N+1)/2 = 5050$.
		(b) The rank of the quantum Fisher matrix and (c) the trace of the quantum Fisher matrix as functions of $\beta$ from $L=6$ (lower dark curves) to $L=12$ (upper light curves).
	}
\end{figure}

It is known that coherent Gibbs states of Ising type models can be represented exactly as an RBM~\cite{gao2017efficient} by associating each edge of the lattice to one hidden unit 
(we provide a self-contained derivation in  Appendix~\ref{sec:ising_to_rbm}).
In particular, the coherent Gibbs state of an Ising-type model defined on a graph $G=(V,E)$ can be described using the RBM with parameters $a=b=0$ and a $|V|$ by $|E|$ sparse weight matrix $w$.

Using this mapping, we construct the quantum Fisher matrix of the RBM representation for coherent Gibbs states .
To sample from the distribution, we have employed the Wolff algorithm~\cite{wolff1989collective} instead of usual local update scheme in this case as it is more efficient close to the transition point.
The spectral profiles of the quantum Fisher matrix for different values of $\beta$ are shown in Fig.~\ref{fig:coherent_ising2d}(a).

The figure shows very similar shape to that of the TFI case when they are deep in the ferromagnetic or paramagnetic phase.
The eigenvalues exhibit a collapsing distribution in the ferromagnetic phase for large $\beta$  and get progressively more  singular as we increase $\beta$. 
Compare this behavior to the TFI for $h < h_c$ depicted in Fig.~\ref{fig:TFI_evals}.
In the paramagnetic phase ($\beta < \beta_c$), we see a stepwise distribution where the step is exactly located at $N(N+1)/2$, very much like the TFI model at large $h$.
Thus for coherent Gibbs states that are deep in each phase, we get the same qualitative behavior of the quantum Fisher matrix in both models.

In contrast to the learned TFI case in Section \ref{sec:results}, the drop-off at $N(N+1)/2$ survives also at criticality. This can be understood by the fact that the quantum Fisher matrix is constructed from the exact coherent Gibbs state which is exactly symmetric in the exchange of spins. Hence the quantum Fisher matrix has zero support on the anti-symmetric subspace also at criticality. 
In  Fig.~\ref{fig:coherent_ising2d}(c), we have plotted the \textit{quantum Fisher information} which is simply the trace of the quantum Fisher matrix for different values of $\beta$. We see that the quantum Fisher information reaches a maximum in the vicinity of the phase transition point, hence acting as an order parameter reminiscent of the magnetic susceptibility. A more detailed analysis of the quantum Fisher information as a witness of phase transitions for this and other models will be presented elsewhere.

\subsection{The XXZ model}\label{sec:XXZ}
We now consider the Heisenberg XXZ model
\begin{align}
	H = \sum_{i=1}^N  \sigma^i_x\sigma^{i+1}_x+\sigma^i_y\sigma^{i+1}_y+\Delta\sigma^i_z\sigma^{i+1}_z.
\end{align}
This model is exactly solvable using the Bethe Ansatz.
The solution shows three distinct phases: (1) a gapped ferromagnetic phase for $\Delta \leq -1.0$, (2) a critical phase for $-1.0 < \Delta \leq 1.0$, and (3) a gapped anti-ferromagnetic phase for $\Delta > 1.0$.
The ground state when $\Delta \leq 1.0$ is a superposition between $\ket{0}^{\otimes N}$ and $\ket{1}^{\otimes N}$.
It is also known that the ground state is in $J_z := \sum_i \sigma_z^{i} = 0$ subspace for $\Delta > -1.0$.
In the critical phase ($-1.0 < \Delta \leq 1.0$), the Hamiltonian is gappless in the thermodynamic limit and the correlation length diverges.
The phase transition at $\Delta=-1.0$ is  first order and an infinite order Kosterlitz-Thouless transition takes place at $\Delta=1.0$.

\begin{figure}
	\includegraphics[width=1.0\linewidth]{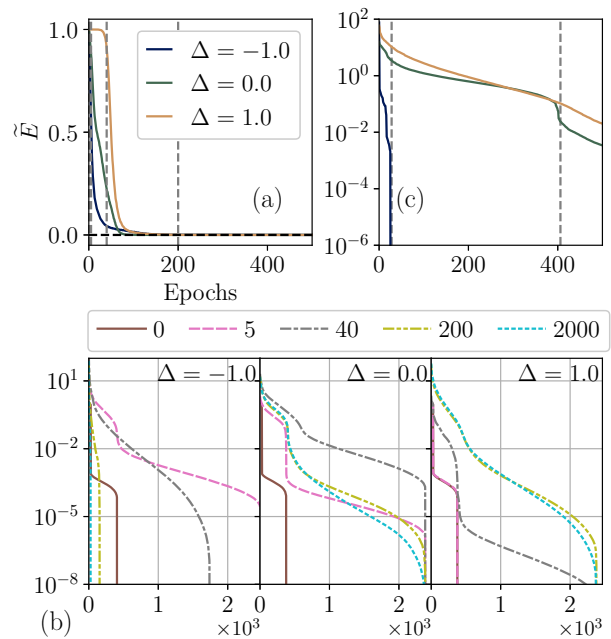}
	\caption{\label{fig:XXZ}
		(a) Rescaled energy as a function of epochs for the XXZ model with $\Delta = -1.0$, $0.0$ and $1.0$ (from darkest to lightest). (b) Spectra of the quantum Fisher matrix at epochs $0$ (solid), $5$ (dashed), $40$ (dot-dashed), $200$ (dot-dot-dashed), and $2000$ (dotted) when $\Delta=-1.0$ (left), $0.0$ (middle) and $1.0$ (right).
	(c) Spectra of converged Fisher matrices. The same colors with (a) are used for $\Delta$. Hyper-parameters $\eta=0.02$ and $\epsilon=0.001$ are used for SR.
	}
\end{figure}

We will again look at the spectral properties of the Fisher information matrix in this model for $\Delta = -1.0, 0.0$, and $1.0$.
For $\Delta= 0.0$ and $1.0$, we have restricted the wave function to the  $U(1)$ symmetric subspace $J_z = 0$ by applying the swap update rule in MCMC.
Figure~\ref{fig:XXZ}(a) shows the convergence of sampled energy over SR iterations. 
We see that SR successfully finds the ground states in all cases but
the initial drift starts later in the XXX case ($\Delta=1.0$).
Slow initial learning when $\Delta=1.0$ is also checked in the spectrum of the quantum Fisher matrix shown in Fig.~\ref{fig:XXZ}(b) where the spectrum begins to change slowly compared to other cases.
We suspect that the ${\rm SU}(2)$ symmetry of the Hamiltonian is related to slow learning in the initial stage.
When we compare the quantum Fisher matrices and the gradient of energies, which are two main ingredients of SR, for different values of $\Delta$, quantum Fisher matrices do not differ much as they only depend on the parameters of the RBM but the gradient of the energy $\nabla_\theta \< H\>$ is much smaller when $\Delta = 1.0$ than other cases.

We  plot the converged spectra  in Fig.~\ref{fig:XXZ}(c). 
Using this, we can extract some information of the converged ground state when $\Delta=-1.0$.
As the first order phase transition occurs at this point, the system has two different types of ground states: one that is a superposition of $\ket{0}^{\otimes N}$ and $\ket{1}^{\otimes N}$ from $\Delta \leq -1.0$ and the other one living in a subspace $J_z=0$ from $\Delta > -1.0$.
As the converged spectrum is singular, we can expect that the ground state found in our simulation is ferromagnetic. We indeed have calculated $\<J_z^2\>$ from Monte-Carlo samples and it gives $\<J_z^2\>/N^2 \approx 0.984$ which means a large portion of the state is in $\ket{0}^{\otimes N}$ and $\ket{1}^{\otimes N}$.
When $\Delta=0.0$ and $1.0$, we see broader converged spectra.
We note that there is a small step at $\sim N(N+1)/2$ when $\Delta=0.0$ even though the whole spectrum is dense.
In comparison, more smooth spectrum is obtained when $\Delta=1.0$.

One should also ask about the behavior of quantum Fisher matrix in the anti-ferromagnetic phase.
However, we found that usual MCMC does not produce unbiased samples in the anti-ferromagnetic phase, so SR does not converge to the real ground state~\footnote{Even though this problem can be solved by applying a local basis transformation that makes the Hamiltonian stoquastic, we did not use such a technique as we want to see how RBM encodes a quantum state without post-manipulation of the problem.}.
As a consequence, we checked the optimization using the exactly constructed quantum Fisher matrix for small enough systems from the probability distribution $|\psi_\theta(x)|^2$.
The result obtained from the exact simulation for the system size $N=20$ is shown in Appendix~\ref{sec:app_xxz_20}.
One observation is that we  see a dense converged spectrum when $\Delta=2.0$ despite the system being gapped. Thus the gap of the system alone does not implies a dense spectrum of the quantum Fisher  matrix.

\section{Implication for optimization}~\label{sec:rmsprop}
In this section, we use the insight gained about the structure of the quantum Fisher matrix to construct a new optimization method for quantum spin systems. The new method allows for significant savings in evaluation time for solving the inverse linear problem in the SR.
Precisely, in each step of SR, we need to solve the linear equation 
\bea Sv = \nabla_\theta \<H\> \eea
for a given quantum Fisher matrix $S$.
Even when the matrix $S$ is well-conditioned, the complexity of solving this equation scales as $O(D^2)$ where $D$ is the dimension of the $S$ matrix, or number of parameters. As $D$ itself scales like $O(\alpha N^2)$, the time cost is quartic in $N$.
This is one of the main reasons why  second order methods, including  natural gradient descent, are not widely used in classical large scale deep  learning applications. 

Our new optimization method can be seen as an extension of RMSProp~\cite{rmsprop}. 
The method provides a significant advantage in computation time as it does not involve solving a large system of linear equations.
However, the method is not always a good approximation of the natural gradient, but rather depends decisively on the  structure of the quantum Fisher matrix.

Before describing our method, we briefly review RMSProp for classical machine learning and how it is related to the Fisher information metric from the viewpoint of Ref.~\cite{martens2014new}.
For convenience, the original RMSProp is described in Appendix \ref{sec:app_rmsprop_para}.
This algorithm improves a naive stochastic gradient descent by using $v_t$, the running average of the squared gradients, to rescale the instantaneous gradient for updating weights.
An observation in Ref.~\cite{martens2014new} is that $v_t$ is a diagonal approximation of the uncentered covariance matrix of gradients when the learning is in the steady state.
When the function we want to optimize $f$ is the logarithmic likelihood (which is typical in classical machine learning), $v_t$ recovers the diagonal part of the Fisher information metric at stationarity. 
The additional square root and $\epsilon$ prefactor in the last step are added  to correct for ``poor conditioning''~\cite{martens2010hessian}.
This provides a plausible argument for why such a simple algorithm works incredibly well.
One can also argue that other popular and efficient optimizers such as Adagard, Adadelta and Adam similarly use a type of diagonal approximation of the Fisher information metric~\cite{martens2014new}.

We now describe our variant of RMSProp applied to the ground state optimization problem.
Using the same principle as above, one may use $\<O\>$ to estimate the diagonal part of the uncentered quantum Fisher matrix $\tilde{S}_{\alpha,\alpha} = \<O_\alpha^\dagger O_\alpha\>$.
The details of the algorithm are outlined in Alg.~\ref{alg:quantum_rmsprop}.
A distinguishing property of this algorithm to the original RMSProp is that it uses different vectors for a gradient decent direction and estimating the curvature: $v_t$ is calculated by $\<O\>$ but the gradient of the energy is used for update in the last step.
The algorithm suggested here is also different from the method used in Refs.~\cite{kessler2019artificial,yang2019deep} that put energy gradient directly to the classical optimizers.

\begin{algorithm}[H]
	\caption{RMSProp for ground state calculation. Hyper-parameters $\beta=0.9$ and $\epsilon=10^{-8}$ are used in our example.}
	\label{alg:quantum_rmsprop}
\begin{algorithmic}[1]
	\Require {$\eta$: Learning rate}
	\Require {$\beta$: Exponential decay rate}
	\Require {$\theta_0$: Initial parameter vector}
	\State {$t \leftarrow 0$ (Initialize timestep)}
	\State {$v_0 \leftarrow 0 $ (Initialize 2nd moment vector)}
	\While {$\theta_t$ is not converged}
		\State {$t \leftarrow t+1$}
		\State {$g_t  \leftarrow$ Gradient of the energy}
		\State {$O_t  \leftarrow \<{O}\>$ }
		\State {$v_t = \beta v_{t-1} + (1-\beta)O_t^* \odot O_t$}
		\State {$\theta_t = \theta_{t-1} - \eta g_t \odot 1/(\sqrt{v_t} + \epsilon)$}
	\EndWhile
\end{algorithmic}
\end{algorithm}

\begin{figure}
	\includegraphics[width=1.0\linewidth]{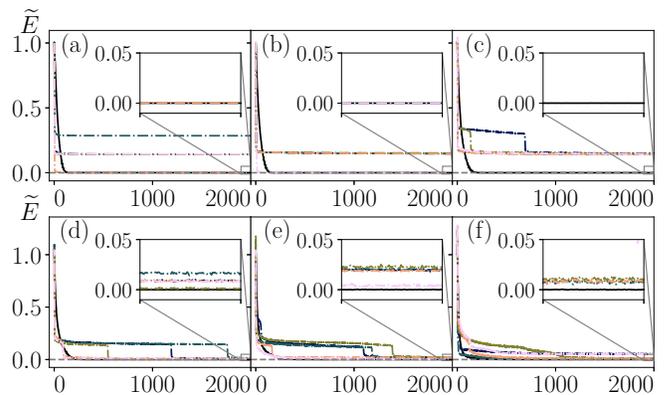}
	\caption{\label{fig:TFI_opt} Epochs versus rescaled energies obtained from the RMSProp (dot-dashed) with different learning rates and the SR with $\eta=0.01$ (black solid). The TFI with the transverse fields from (a) $h=0.0$ to (f) $1.0$ are used. For the RMSProp, we used learning rates $\eta = 1.4 \times 10^{-3}$ (the darkest) to $2.2 \times 10^{-3}$ (the lightest) with the interval $0.2 \times 10^{-3}$.}
\end{figure}

We have tested the proposed version of RMSProp using different learning rates $\eta$ for the TFI.
The results for the ferromagnetic phase and the critical case ($h=0.0$ to $1.0$) are shown in Fig.~\ref{fig:TFI_opt}. 
For small $h$, we see that RMSProp gets easily stuck in local minima unlike SR. 
When $h=0.0$ and $0.2$, the figure shows that the energy converges to that of the ground state for some learning rate $\eta$.
However, such a convergence is probabilistic.
For $h=0.0$, $0.2$ and $0.4$, we ran the same simulation several times and found that, for any $\eta$, some instances converge to the ground state whereas others get stuck in local minima.
In contrast, SR works properly for a wide range of hyper-parameters and $h$, for which the energy converges to the ground state regardless of the choice of the learning rates $\eta=[0.005, 0.01, 0.02]$.

For larger $h$ such as $h=0.6, 0.8$, the proposed RMSProp shows better convergence behaviors for most values of $\eta$ but it still show stepwise dynamics. 
In the critical case $h=1.0$, the learning curves of RMSProp are smooth and insensitive to the choice of the learning rate, suggesting that the system no longer gets stuck in problematic local minima. 

Our results suggest that preserving the singular nature of the quantum Fisher matrix is essential for ensuring convergence to the ground state energy. 
Indeed, the converged quantum Fisher matrices studied in Appendix~\ref{sec:more_fisher} show that the diagonal of the Fisher matrices give rank $N+M=112$ for $h=0.0$ and full rank ($NM+N+M = 2464$) for other values of $h$.
In contrast, the real ranks of the quantum Fisher matrices (measured by counting the number of eigenvalues larger than $10^{-10}$) are given as $1, 78, 242, 726, 1698, 2464$ for $h=0.0, 0.2, 0.4, 0.6, 0.8$ and $1.0$, respectively.

We still note that even though the rank provides a plausible argument for the behavior of the learning curves, it does not for the converged energies; the converged energies for $h=0.8$ and $1.0$ are slightly larger than the ground state energies.
Moreover, the convergence behavior in the paramagnetic phase ($h > 1.0$) is more complicated and cannot be solely explained from the quantum Fisher matrix. A partial reason is that the path taken by RMSProp deviates from that of the SR in initial stage of learning (see Appendix~\ref{sec:app_rmsprop_para}). Detailed investigations in this regime remain for future work.

\section{Conclusion}

We have initiated a detailed study of the quantum information geometry of learning ground states of spin chains in the artificial neural neural network framework. We have focused on  complex restricted Boltzmann states and the stochastic reconfiguration method which implements a quantum version of Amari's natural gradient update scheme. Our main result is that the eigenvalues and eigenvectors of the quantum Fisher matrix reflect both the learning dynamics, which is unsurprising, as well as the intrinsic static phase information of the model under study, which is rather surprising. In particular, we found that in the entire non-critical ferromagnetic phase of a number of models, the spectrum of the quantum Fisher matrix has reduced rank. The matrix becomes highly singular in regions of the phase that are close to product states. In critical phases, the spectrum becomes smooth with more and more eigenvectors contributing to the information geometry landscape. 

We have  identified a universal behavior of the leading eigenvectors of the quantum Fisher matrix: they all  convey  little entanglement, as measured by the entanglement entropy between the visible and hidden layers. This, in combination with the insight that critical models have smooth spectra, suggests that correlations in complex RBM Ansatz are preferentially represented in the bulk of the information geometry space. Our interpretation of this key dynamical feature of RBM learning is that the model preferentially chooses stable representations, where the entropy of the landscape dominates over the energy. A similar phenomenon is classical supervised machine learning is frequently observed in discussion of ``generalization''.  
Finally, we explored strategies for diagonal approximations of the quantum Fisher matrix, and found that their success crucially depends on the phase of the model under study. We therefore do not expect any diagonal approximation of the quantum Fisher matrix to be effective in general.

\section{Acknowledgements} We thank S. Trebst and D. Gross for helpful discussions. We acknowledge support from the DFG (CRC TR 183), and the ML4Q excellence cluster. 
Source codes for the current manuscript can be found in CYP's github repository~\cite{CYPgithub}.
The numerical simulations were performed on the CHEOPS cluster at RRZK Cologne.

\begin{appendix}
\section{Stochastic reconfiguration}\label{sec:app_sr}
For the readers convenience, we derive the stochastic reconfiguration method of Sorella~\cite{sorella1998green,sorella2001generalized}. 
The main idea of Stochastic Reconfiguration (SR) is to modify the parameters of a trial wave function in such a way that it approaches the ground state along a path dictated by the projection $\1 - \epsilon H$, where $\epsilon$ is chosen such that  $\1 - \epsilon H\geq0$.

Let $\ket{\psi_\theta}$ be a state in our ansatz class, with $\theta$ its vector of parameters. From now on, we will suppress the parameters $\theta$. Then, for sufficiently small $\epsilon$, we can write
\be (\1-\epsilon H)\ket{\psi}=e_0\ket{\psi}+\sum_\alpha e_\alpha \ket{\psi_\alpha}+\ket{\psi^\perp},\label{eqn:SR1}\ee
where $\ket{\psi_\alpha}=\frac{\partial}{\partial \theta_\alpha}\ket{\psi}$, $\{e_\alpha\}$ are coefficients, and $\ket{\psi^\perp}$ is a state in the orthogonal subspace.
Note the identity $\ket{\psi_\alpha}=O_\alpha\ket{\psi}$, where the operators $O_\alpha$ are defined as:
\be O_\alpha\ket{x}=\frac{\partial \log (\langle x| \psi\rangle)}{\partial \theta_\alpha}\ket{x},\ee
were $\ket{x}$ is the computational basis. 

We can now obtain a system of linear equations for the $e_\alpha$ coefficients by multiplying Eqn. (\ref{eqn:SR1}) by $\bra{\psi}$ and by $\bra{\psi_\alpha}$ to get
\bea 1-\epsilon \avr{H}&=&e_0+\sum_\alpha e_\alpha \avr{O_\alpha}\\
\avr{O^\dag_\alpha}-\epsilon \avr{O^\dag_\alpha H}&=&e_0\avr{O^\dag_\alpha}+\sum_\beta e_\beta\avr{O^\dag_\alpha O_\beta}\eea

The averages are taken in the states $\ket{\psi}$. We can then solve for $e_0$ to get 
\be \sum_\beta S_{\alpha,\beta}e_\beta=-\epsilon R_\alpha,\ee
where the matrix $S$ is given by 
\be S_{\alpha,\beta}=\avr{O^\dag_\alpha O_\beta}-\avr{O^\dag_\alpha}\avr{O_\beta},\ee
and the vector $R_\alpha$ is given by 
\be R_\alpha=\avr{O^\dag_\alpha H}-\avr{O^\dag_\alpha}\avr{H}.\ee

We can now identify the coefficients $e_\alpha$ as the update coefficients for the variables $\theta_\alpha$, up to an overall constant $e_0$, which can be interpreted as the learning rate. The SR update scheme can then be summarized as:
\be \theta_\alpha\rightarrow \theta_\alpha-\eta \sum_\beta (S+\epsilon \1)^{-1}_{\alpha,\beta}R_\beta,\ee
for some learning rate $\eta$. Here, $\epsilon$ is regularization constant that is typically $\sim 10^{-3}$.

\begin{table}[H]
\centering
\bgroup
\def\arraystretch{1.5}% 
\begin{tabular}{| c || c | c | c | c|} 
 \hline
	Model & Lattice size & Monte-Carlo update & $\eta$ & $\epsilon$ \\ [0.5ex] 
	\hline\hline
	1D TFI & $28$ & Spin flip & 0.01 & 0.001 \\
	\hline
	2D TFI & $5\times5$ & Spin flip & 0.002 & 0.001 \\
	\hline
	XXZ & $28$ & Swap & 0.02 & 0.001 \\
	\hline
\end{tabular}
\egroup
\caption{\label{tab:numerics} Parameters used for the simulations in the main text.} 
\end{table}

\section{Numerics} \label{sec:app_numeric}
For numerical simulation, we set the ratio between the numbers of hidden units and visible units of the complex RBM to $\alpha = M/N = 3$.
Thus the RBM has  $(\alpha+1)N + \alpha N^2$ parameters overall ($N$ and $\alpha N$ for biases and $\alpha N^2$ for the weight matrix $w$).
To sample from the RBM, Markov chain Monte-Carlo (MCMC) method enhanced with parallel tempering was employed~\cite{choo2018}.
We  used 16 parallel Markov chains with linearly divided temperatures from $1/16$ to $1$. 
For each Markov chain, we used local spin flip updates for the transverse field Ising models (1D and 2D) and total magnetization conserving swap updates for the XXZ model.
To directly compare the results from variational Monte-Carlo with exact diagonalization, we have used the size of system $N=28$ for 1D models, $L \times L$ with $L=5$ for 2D TFI, and imposed the periodic boundary condition.
In our case, SR has two hyper-parameters: the learning rate [$\eta$ in Eq.~\eqref{eq:SR_update}] and the regularization $\epsilon$. 
These hyper-parameters in our simulation results are summarized in Table~\ref{tab:numerics}.

\section{quantum Fisher matrix of random RBM} \label{sec:random_RBM}
We provide an explanation of the stepwise structure of the spectrum of the quantum Fisher matrix upon small random initialization of the weights.
The quantum Fisher matrix is broken up into three main sectors: $[a,b,w]$, corresponding to the visible biases, the hidden biases and the weights.

As in the main text, we use $N = |a|$ and $M = |b|$ to indicate the number of visible and hidden units, respectively.
In our simulations, the weights are initialized to be Gaussian distributed with an average magnitude of order $\sigma = 10^{-2}$. We therefore make the following assumption about the initial state: \textit{the classical probability distribution associated with the initial quantum state is close to the identity, and in particular is separable}. This implies that each spin has zero expectation value at initialization $\avr{x_j}=0$ for all $j$, and that $\avr{x_jx_k}\propto \delta_{jk}$ for all $j k$. 

As the entries of the visible biases block are: 
\be S_{a_i,a_j}=\avr{x_i  x_j  } - \avr{x_i} \avr{x_j}=\delta_{ij}.\ee
we get the identity matrix for the $a$ part. 
The covariance between the visible and hidden units involves the term $\<x_i \tanh (\chi_j (x))\>$.
Recall that the argument of the hyperbolic tangents are 
\be \chi_j(x) = b_j +\sum_i w_{ij} x_i.\ee
where $b_j$ are the hidden biases and $w_{ij}$ are the weights connecting the hidden and visible units. 
Under the assumption that all parameters are small, we approximate $\tanh(\chi_j(x)) \approx \chi_j(x) $. 
Then 
\begin{align}
	\<x_i \tanh (\chi_j(x))\> &\approx \<x_i \chi_j(x)\> = b_j\<x_i\> + \sum_k w_{kj} \<x_i x_k\> \nonumber \\
																&\approx \sum_k w_{k j}  \delta_{ik} = w_{i j}.
\end{align}
Likewise, we can obtain the full unary part ($[a,b]$) of the $S$ matrix as
\begin{align}
	S_{\rm un} = \begin{pmatrix}
		\1_{N} & w \\
		w^\dagger & w^\dagger w
	\end{pmatrix}.
\end{align}
We can easily see this is rank $N$ as the first $N$ row generates the remaining rows.
This explains the first $N$ eigenvalues which are $O(1)$. 

Next, the $w$ part of the quantum Fisher matrix is  given by 
\begin{align} 
	(S_{w})_{i j, i'j'}	&=\avr{x_i \tanh(\chi_j(x))^* x_{i'} \tanh(\chi_{j'}(x)) } \\
								&\quad - \avr{x_i \tanh(\chi_j(x))^*}\avr{ x_{i'} \tanh(\chi_{j'}(x)) } ,
\end{align}
where $i,i'$ label the visible units and $j,j'$ label the hidden units. 
Using the expansion
\begin{align}
	&\avr{x_i \tanh(\chi_j(x))^* x_{i'} \tanh(\chi_{j'}(x)) } \nonumber \\
	&\qquad \approx b_j^* b_{j'} \delta_{ii'} + \sum_{k k'} w_{k i'}^* w_{k' j'} \<x_i x_k x_j x_{k'}\>,
\end{align}
we have  
\be S_w(b)=(\1\otimes w^\dag) X(\1 \otimes w) + \1_{n}\otimes |b\rangle \langle b|,\label{eq:S_W}\ee
where $w$ is the $N\times M$ matrix of weights, $|b\> = \sum _j b_j |j \rangle$ is a vector form of the bias $b$,
and 
$X=\sum_{ijkl} x_{ikjl} \ket{ik}\bra{jl}$ with $x_{ikjl}=\avr{x_ix_kx_jx_l} - \< x_i x_k \> \< x_j x_l\>$. 
Using the assumption of small initial weights, we have
\be x_{ikjl} = \delta_{ij}\delta_{kl}+\delta_{il}\delta_{jk} - 2\delta_{ikjl}.\ee
Then the $X$ matrix is approximately 
\begin{align}
		X &= \sum_{jk} (\ket{jk}\bra{jk}+\ket{jk}\bra{kj}) - 2 \sum_j \ket{jj}\bra{jj}\nonumber \\
&= \1 + V - 2 \sum_j \ket{jj} \bra{jj}
\end{align}
where $V = \sum_{jk} \ket{jk}\bra{kj}$ is the swap operator.
The rank of $X$ is given by $N(N-1)/2$. Moreover, $X$ is the projector that preserves the symmetric states except the copied state, i.e. $X (\ket{ab} + \ket{ba}) \propto \ket{ab} + \ket{ba}$ when $a \neq b$ but $X \ket{aa} = 0$.

When $b=0$, the whole covariance matrix is given by $S = S_{\rm un} \oplus S_{w}$ and the matrix $S_w$ [Eq.~\eqref{eq:S_W}] has rank $N(N-1)/2$. This explains the small sub-leading eigenvalues of  order $O(\sigma^2)$.

However, the block-diagonal assumption breaks down when we have non-zero bias in the hidden layer ($b\neq 0$) as we have off-diagonal blocks between the unary and $w$ part.
An additional $\1\otimes \ket{b}\bra{b}$ also enters into $S_w$.
Still, it is not difficult to see that this does not change the overall rank. 
A precise calculation gives 
\begin{align}
	S(b) = \left(\begin{array}{ccc}
			\1_{N} & w & \1\otimes \< b| \\
			w^\dagger & w^\dagger w & w \otimes \< b | \\
			| b\rangle \otimes \1 & | b\rangle \otimes w & S_{w}(0) + \1\otimes |b\rangle \langle b|
		\end{array}\right)
\end{align}
up to third order corrections.
It is simple to see that first $N$ rows still generate the next $M$ rows. 
Moreover, applying $|b\rangle$ to the first $N$ rows gives the additional terms in the last $NM$ rows so the rank of the $S$ matrix from the $w$ part also does not change.
Thus we have exactly the same rank even when we turn on hidden biases $b$.

\begin{figure*}
	\centering
	\includegraphics[width=0.90\textwidth]{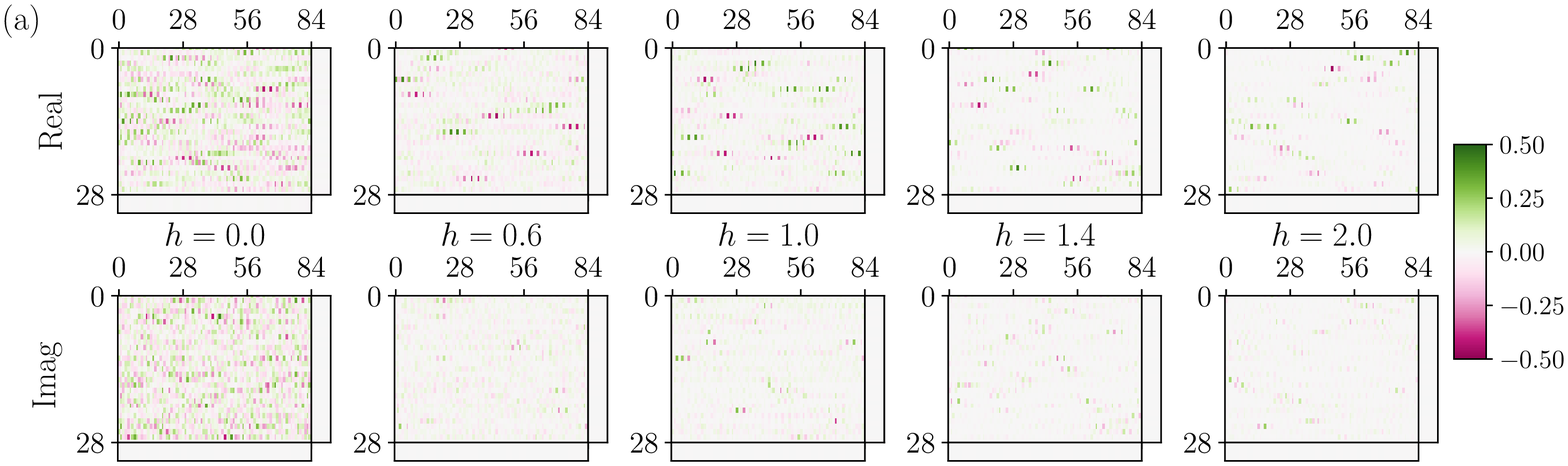}
	\includegraphics[width=0.90\textwidth]{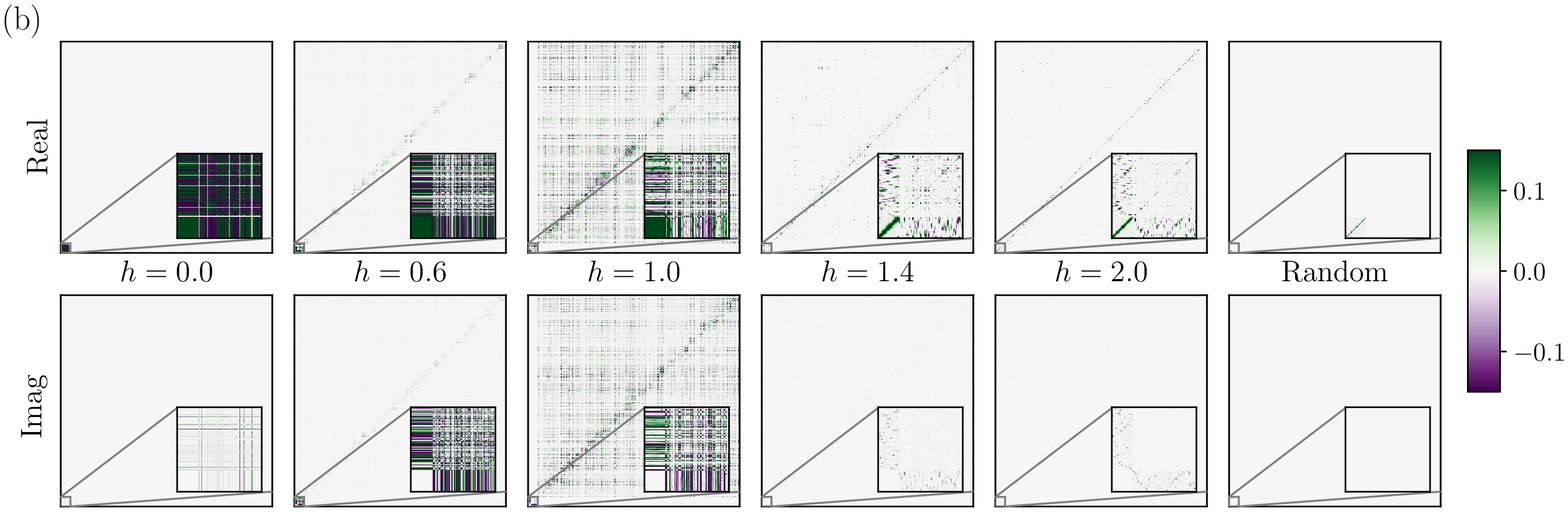}
	\caption{\label{fig:TFI_Smat}
		(a) Converges weights $(a, b, w)$ for the TFI model with different values of $h$. The large rectangle shows the weights $w$, whereas the small strips show the biases $a$ and $b$, which are much weaker in magnitude than the leading weights. 
		(b) Real and imaginary parts of the quantum Fisher matrix after convergence for the TFI as well as randomly initialized RBM. Insets show the correlation between unary variables.
	The whole matrix is order $N+M+NM = 2464$ and the unary part is order $N+M = 112$. The covariance between visible units are small left bottom corner of the unary part.
	}
\end{figure*}

\section{Further properties of the quantum Fisher matrix}~\label{sec:more_fisher}
In this section, we investigate further properties of the quantum Fisher matrix.
We use the same numerical data as in the main text; the TFI with  system size $N=28$.

\subsection{Converged weights}
Converged parameters of neural networks are often claimed to reveal features of the data or system under study ~\cite{carleo2017solving,sehayek2019learnability}.
We compare the converged weights and the quantum Fisher matrix for different values of $h$ in Fig.~\ref{fig:TFI_Smat}.
We find that, in contrast with the spectral information of the quantum Fisher matrix, it is difficult to infer any information from the converged weights of the network. 
For example, converged weights for $h=0.6, 1.0$ and $1.4$ are not sensibly different, whereas the quantum Fisher matrices reveal essential features of the phase of the system.

This brings to light one the of the key subtleties of RBM Ans\"{a}tze, which is the extreme redundancy of representation.
Let us illustrate this fact by constructing  three completely different solutions of the RBM parameters that (approximately) represent the same quantum state $|0\>^{\otimes N} + |1\>^{\otimes N}$.
As a first solution, consider the one obtained from our numerical simulation Fig.~\ref{fig:TFI_Smat} (a). 
This solution is fully complex, i.e. real and imaginary parts of the weights are both non-zero.
On the other hand, a real solution can be found from the coherent Gibbs states for classical Ising model as discussed in Appendix~\ref{sec:ising_to_rbm}. The  state is obtained by letting $J_{ij}=-1$ and $\beta \rightarrow \infty$ for a classical Ising model defined on any graph that does not have an isolated vertex.
We note that the parameters obtained using this scheme are real as $e^{-\beta J_{i,j}} \geq 1$ (see Appendix~\ref{sec:ising_to_rbm} for details).
Finally, it is also possible to represent this state only using pure imaginary parameters.
By letting $a=0$, $b = (i \pi/2, \cdots, i \pi /2)$, and the weight $w$ as
\begin{align}
	w_{i, j} = \begin{cases}
		i \pi/4, & \text{if } j = i+1 \\
		0,& \text{otherwise}
	\end{cases}.
\end{align}
It is clear from these examples that inferring information of quantum states solely from the activation parameters of the RBM is very ambiguous.

\subsection{Non-zero elements of Fisher information matrix}
We investigate the rank of the quantum Fisher matrix more closely. 
Let us first focus on the ferromagnetic phase ($h < 1.0$).
In the main text, we have shown that the rank of the quantum Fisher matrix increases as $h$ increases. 
A question we are interested in is how non-zero elements are distributed in unary and $w$ parts of the matrix.
To answer this question, we use the quantum Fisher matrix itself after convergence plotted in Fig.~\ref{fig:TFI_Smat}(b).
When $h=0.0$, we see that the Fisher information matrix only has non-zero elements in the unary part.
In contrast, the $w$ part of the matrix shows non-zero elements (especially in diagonal part) when $h=0.6$.
To see this clearly, we have counted the number of diagonal elements of the quantum Fisher matrix that are larger than $10^{-4}$.
It shows there are $N+M=112$ such diagonal elements when $h=0.0$ but $N+M+NM=2464$ for all larger $h=0.2, 0.4, 0.6, 0.8$.
As the rank of the full matrix is small even for larger $h$, the non-zero elements in the $w$ part in this case implies the eigenvectors with dominant eigenvalues have compelling $w$ part.
In addition, this provides an argument why RMSProp that is studied in Sec.~\ref{sec:rmsprop} works badly for small $h$.

Next, we consider the paramagnetic phase ($h > 1.0$).
In the main text, we have shown that the Fisher information matrix when $h=2.0$ shows a step at $N(N+1)/2$.
The whole shape of the spectrum remains similar for smaller $h$ even though the location of step can be little shifted.
Compared to the randomly initialized RBM, we see larger diagonal elements in $w$ part. 
As Fig.~\ref{fig:TFI_evals} shows that eigenvalues between $N$th to $N(N+1)/2$ are much larger for the converged Fisher information matrix than the random RBM, we expect that $w$ part of the matrix contributes to these eigenvalues.
To test this, we have diagonalized only the $w$ part of quantum Fisher matrix when $h=2.0$ where we could observe a step at $N(N-1)/2$. 
Thus despite the whole spectrum does not show a clear step at $N$-th eigenvalue, we may still consider that $N$ eigenvalues are from the unary part and $N(N-1)/2$ are from the $w$ part.
We also found that all diagonal elements of the quantum Fisher matrix is larger than $10^{-2}$ when $h\geq 1.0$ so the diagonal approximation of the quantum Fisher matrix is full rank.

\begin{figure}[t]
	\centering
	\includegraphics[width=1.0\linewidth]{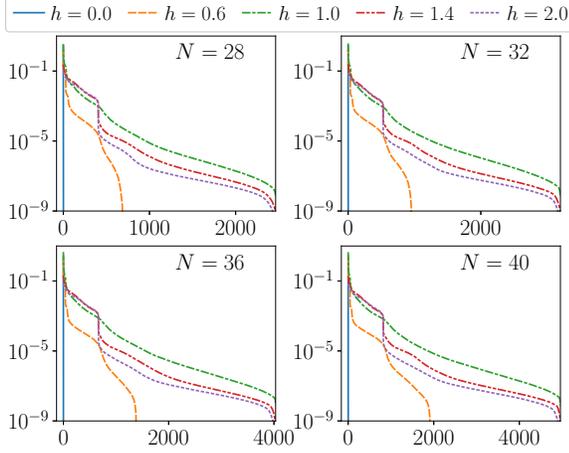}
	\caption{\label{fig:TFI_varing_N}
		Normalized eigenvalues $\lambda_i/N$ of the converged quantum Fisher matrix for the TFI with the system sizes $N=28$ to $40$ (from (a) to (d)). The transverse fields $h=0.0$ (solid), $0.6$ (dashed), $1.0$ (dot-dashed), $1.4$ (dot-dot-dashed), and $2.0$ (dotted) are used. The shapes of the distributions are independent to $N$.
	}
\end{figure}

\subsection{System size dependence of the spectral profile}
When we use the same parameter $\alpha = M/N$ and the Hamiltonian, we observe that spectra of the converged Fisher information matrix behaves almost the same for varying $N$. 
In Fig.~\ref{fig:TFI_varing_N}, we show the spectra of the converged quantum Fisher matrix for different values of $N=[28,32,36,40]$ using the TFI with different values of $h=[0.0, 0.6, 1.0, 1.4, 2.0]$. 
We clearly see that eigenvalue distributions for the same $h$ vary only little with the change of the system size $N$.
Still, it is not easy to make an exact correspondence between the results from different $N$ as the order of the quantum Fisher matrix is given by $\alpha N^2 + (\alpha+1)N$ which is not monomial. Thus there is no single constant scale factor we can use for rescaling the results. Still, this suggests that the spectrum of the quantum Fisher matrix can be used as a faithful  diagnostic tool on small systems to infer qualitative behavior on larger systems.

\section{Coherent Gibbs states for classical Ising models} \label{sec:ising_to_rbm}
We consider a classical Ising model defined on a graph $G=(V,E)$ where $V = \{i\}$ is the set of vertices and $E = \{(i, j)\}$ is the set of edges. 
We assign binary values $x_i = 1$ or $-1$ to each vertex and interaction strengths $J_{i,j}\in \mathbb{R}$ to each edge $e = (i,j) \in E$.
The Hamiltonian of this model is given by
\begin{align}
	H(x) = \sum_{(i,j)\in E} J_{i,j} x_i x_j.
\end{align}

Then our objective is finding parameters of the RBM $[a,b,w]$ that describe coherent Gibbs states for the given $\beta$, i.e. solving the equations
\begin{align}
	\psi_\theta(x) = e^{a \cdot x} \prod_{j=1}^M 2 \cosh \chi_j(x) = c \exp[- \beta H(x)/2] \label{eq:ising_to_rbm}
\end{align}
for all $x=\{-1,1\}^N$. Here, $\chi_j(x) = \sum_i w_{ij} x_i + b_j$ and $c$ is a constant that can be freely chosen as our RBM does not use a specific normalization.

As the $H(x)$ is symmetric under overall flip ($x \rightarrow -x$), we first consider $\mathbb{Z}_2$ symmetric RBM that has zero biases, i.e. $a=b=0$.
Then we can simplify the equation to
\begin{align}
	\prod_{j=1}^M 2 \cosh (\sum_k w_{k j} x_k) = c \prod_{ (i,j) \in E}\exp[-\beta J_{i,j} x_i x_j/2].
\end{align}
We can find such a $w$ easily by letting $M=|E|$ and equating each term using a column of $w$ in the left hand side to the term in the right hand side using an edge.
In other words, we solve
\begin{align}
	2 \cosh( \sum_k w_{k e} x_k) = c_e \exp[-\beta J_{i,j} x_i x_j/2] \label{eq:ising_to_rbm_term}
\end{align}
for all $e \in E$ where $c_e$ is a constant assigned to each edge $e$ that gives $c=\prod_{e \in E} c_e$. 
Setting all $w_{k e}=0$ if $k\neq i,j$, we then need to solve the coupled equations
\begin{align}
		2 \cosh(w_{ie}+w_{j e})&= c_e e^{-\beta J_e/2} \\
		2 \cosh(w_{ie}-w_{j e})&= c_e e^{\beta J_e/2}
\end{align}
These equations can be solved for any $\beta J_{i,j}$ as $w$ is a complex matrix. 

For the two dimensional Ising model we consider in the main text, $J_{i,j} = -1$ for all edges $(i,j) \in E$ that connect any neighboring vertices in 2D lattice. In this case, we can easily get a real solution $w_{ie} = w_{je} = \cosh^{-1}[e^{\beta}]/2$.

\begin{figure}
	\centering
	\includegraphics[width=0.98\linewidth]{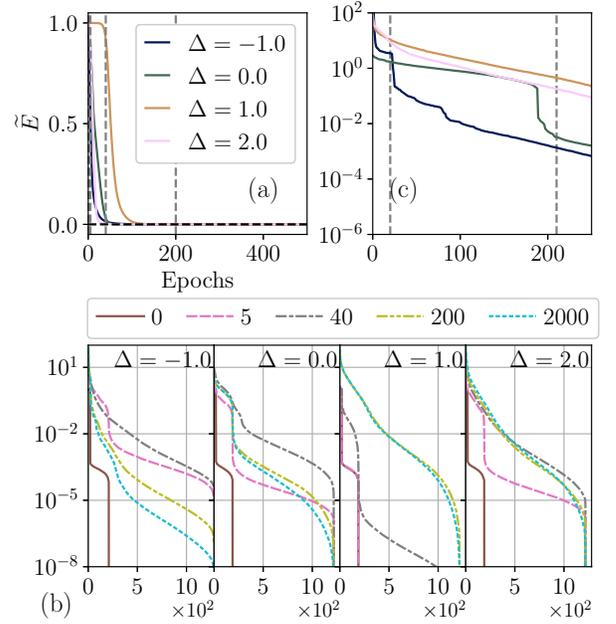}
	\caption{ \label{fig:XXZN20}
		Numerical results of the XXZ model with size $N=20$ using exactly constructed wave functions. (a) Normalized energy $\widetilde{E} = (\<E\>-E_{\rm ed})/(E_0 - E_{\rm ed})$ as a function of epochs. The interaction strengths from $\Delta = -1.0$ (the darkest) to $2.0$ (the lightest) are used.
		(b) Dynamics of the spectrum of the Fisher information matrix at epochs $0$ (solid), $5$ (dashed), $40$ (dot-dashed), $200$ (dot-dot-dashed), and $2000$ (dotted). Interaction strengths from $\Delta=-1.0$ (the leftmost) to $2.0$ (the rightmost) with the interval $1.0$ are used.
		(c) Spectrum of converged Fisher information matrix. The same colors with (a) are used to indicate $\Delta$.
	}
\end{figure}

\section{The XXZ model using exact wave functions}\label{sec:app_xxz_20}
In the main text, we studied the Heisenberg XXZ model using variational quantum Monte-Carlo. 
There the observables such as the quantum Fisher matrix and the energy gradient are calculated from the samples obtained from MCMC.
In this section, we study the same system using exactly constructed wave functions instead of MCMC.
A modified step of each iteration of SR is as follows. 
First, we calculate all components of the wave function $\psi_\theta(x) = e^{a \cdot x} \prod_j 2 \cosh \chi_j$ in the computational basis. 
Then we obtain the normalization factor by calculating the exponential sum $Z=\sum_{\{x\}} |\psi_\theta(x)|^2$.
Using this result, the energy gradient and the Fisher information matrix are also calculated by computing Eqs.~(\ref{eq:S_elt},\ref{eq:Egrad}) exactly and parameters are updated accordingly.
As we do not sample from the distribution, the algorithm is not stochastic anymore. 
Thus we would call this method exact reconfiguration (ER) instead of SR.
We note that ER is extremely expensive in computation since we need to calculate several exponential sums for each iteration.

\begin{figure}[t]
	\centering
	\includegraphics[width=0.95\linewidth]{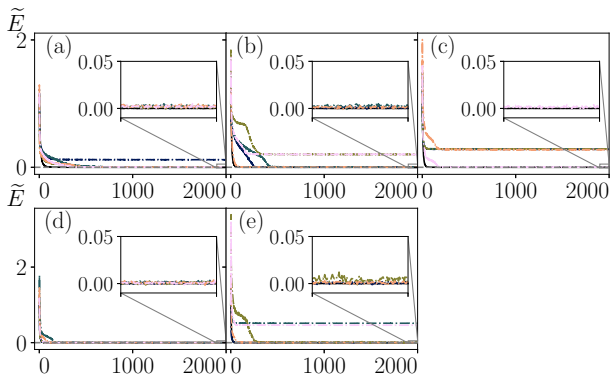}
	\caption{\label{fig:rmsprop_para}
		Rescaled energy $\widetilde{E}$ as a function of epochs for TFI in the paramagnetic phase using the RMSProp (dot-dashed) and the SR with the learning rate $\eta=0.01$ (black solid). Results from the transverse field (a) $h=1.2$ to (e) $2.0$ are shown. Learning rates $1.4\times10^{-3}$ (the darkest) to $2.2 \times 10^{-3}$ (the lightest) are used for the RMSProp. 
	}
\end{figure}

Using ER, we have simulated the XXZ model with the system size $N=20$ that is tractable using current CPUs.
The result is shown in Fig.~\ref{fig:XXZN20}. 
There are two noteworthy features: 
First, the converged spectrum when $\Delta=-1$ shows a broader spectrum as compared to  Fig.~\ref{fig:XXZ} in the main text.
We conjecture that this is related to the fact that the ground state found using ER has more component in $J_z=0$ subspace compared to SR case.
Indeed, we have $\<J_z^2\>/N^2 \approx 0.963$ which is slightly smaller than what is found in the SR case in the main text.
Second, the converged quantum Fisher matrix shows a smooth spectrum when $\Delta=2.0$ even though the system has a gapped anti-ferromagnetic ground state. It implies that a smooth spectrum of the converged quantum Fisher matrix is not sufficient to infer criticality.

\vspace*{0.3em}

\section{RMSProp in the paramagnetic phase}\label{sec:app_rmsprop_para}
We study in this Appendix RMSProp introduced in Sec.~\ref{sec:rmsprop} for the paramagnetic phase of TFI.
The learning curves for $5$ different values of $h$ are shown in  Fig.~\ref{fig:rmsprop_para}.
We can see that the learning curves are more complex than those from the ferromagnetic and the critical cases.
Specifically, we have three distinct observations as follows.
First, there is a spike of the rescaled energy that goes up in the initial stage of learning. In addition, the size of the spike grows with $h$.
This means that an initial direction that optimizer selects is much far from the optimal direction.
Second, the properties of the quantum Fisher matrix are not much relevant to the learning dynamics of the RMSProp.
In Appendix~\ref{sec:more_fisher}, we have shown that the properties of the quantum Fisher matrix do not change much within the paramagnetic phase.
However, the learning curves from the RMSProp do not show similarity between different values of $h$.
Third, the converged energy can be as low as that of the SR case. 
This is interesting as the optimizer sometimes finds the proper solution even though the learning dynamic shows poor behavior. 

From these observations, we suspect that RMSProp takes a different learning pathway than SR in the paramagnetic phase.
To understand the applicability and details of the learning dynamics of the algorithm better, more detailed investigations such as tracking the path of optimization are required.
We leave such a detailed investigation of this optimizers and the comparison to other optimizers for future work. 

\begin{algorithm}[H]
	\caption{RMSProp. Here, $\odot$ is the element-wise product of two vectors.}
	\label{alg:classcial_rmsprop}
\begin{algorithmic}[1]
	\Require {$\eta$: Learning rate}
	\Require {$\beta$: Exponential decay rate}
	\Require {$\theta_0$: Initial parameter vector}
	\State {$t \leftarrow 0$ (Initialize timestep)}
	\State {$v_0 \leftarrow 0 $ (Initialize 2nd moment vector)}
	\While {$\theta_t$ is not converged}
		\State {$t \leftarrow t+1$}
		\State {$g_t  \leftarrow \< \nabla_{\theta} f(\theta_{t-1})\>$}
		\State {$v_t = \beta v_{t-1} + (1-\beta)g_t \odot g_t$}
		\State {$\theta_t = \theta_{t-1} - \eta g_t \odot 1/(\sqrt{v_t} + \epsilon)$}
	\EndWhile
\end{algorithmic}
\end{algorithm}

\end{appendix}

%

%%%%%%%%%%%%%%%%%%%%%%%%%%%%%%%%%%%%%%%%%%%%%%%%%%%%%%%%%%

\end{document}